\documentclass[twocolumn,showpacs,preprintnumbers,amsmath,amssymb]{revtex4}

\usepackage{graphicx}
\usepackage{dcolumn}
\usepackage{bm}

\begin{document}

\title{Decoherence of two qubits in a non-Markovian squeezed reservoir}

\author{Fa-Qiang Wang}
\author{Zhi-Ming Zhang}
\email{zmzhang@scnu.edu.cn}
\author{Rui-Sheng Liang}%
\affiliation{Lab of Photonic Information Technology, School of Information and Photoelectronic Science and Engineering, South China Normal University, Guangzhou 510006, China}%

\date{\today}

\begin{abstract}
The decoherence of two initially entangled qubits in a squeezed
vacuum cavity has been investigated exactly. The results show that,
first, in principle, the disentanglement time decreases with the
increasing of squeeze parameter $r$, due to the augmenting of
average photon number of every mode in squeezed vacuum cavity.
Second, there are entanglement revivals after complete
disentanglement for the case of even parity initial Bell state,
while there are entanglement decrease and revival before complete
disentanglement for the case of odd parity initial Bell state. The
results are quite different from that of the case for qubits in a
 vacuum cavity.

\end{abstract}

\pacs{03.67.Mn,03.65.Yz,03.65.Ud,42.70.Qs}
\maketitle
\section{Introduction}
\par In recent years, the entanglement dynamics of qubits, coupled with environment, has
attracted much attention. Many works have been devoted to the
phenomenon termed as ``entanglement sudden
death"(ESD)\cite{dio,yu1,yu2,mp,mf,ikr,hu,asm,sch,daj,cao,bel,xu,jing,hou}.
It is shown that spontaneous disentanglement, for two-level atom
model, may take only a finite-time to be completed, while local
decoherence (the normal single-atom transverse and longitudinal
decay) takes an infinite time\cite{yu2}. And, for a non-Markovian
reservoir of initially vacuum cavity, there is revival phenomenon,
which allows the two-qubit entanglement to reappear after a dark
period of time, during which the concurrence is zero\cite{bel}. In
general, the characteristic of the environment plays an important
role in the evolution of multi-particle entanglement. Up to now,
many typical environments have been investigated, such as, vacuum,
squeezed vacuum, multimode vacuum cavity, single mode cavity and so
on.
\par Recently, cavity systems with very strong couplings have been discussed\cite{mei}.
 Generally, in atom-field  cavity  systems, this ratio is typically of the order  $10^{-7}\sim10^{-6}$. However, the ratio may
 become order of magnitudes larger in solid state systems, and the full Hamiltonian, including the  virtual processes
 (counter-rotating terms), must be considered\cite{iri}.
 In this paper, we will focus on the decoherence of two qubits strongly coupled with
 a non-markovian squeezed vacuum reservoir by the method in Ref.\cite{Ish}.
  In section.2, the reduced non-perturbative non-Markovian
 quantum master equation of atom is derived and its exact solution is obtained.
 In section.3, the decoherence of two initially entangled atoms, coupled with two cavities separately,
 has been discussed. The conclusion is given in section.4.

 \section{\label{sec:model} Model and exact solution}
 \subsection{Hamiltonian and non-perturbative master equation}
\par Now we restrict our attention to two noninteracting two-level
atoms A and B coupled individually to the cavity field. To this aim,
we first consider the Hamiltonian of the subsystem of a single qubit
coupled to its reservoir as
\begin{equation}\label{e1}
    H=H_{a}+H_{r}+H_{ar}
\end{equation}
where
\begin{eqnarray}
  H_{a} &=& \omega_{0}\frac{\sigma_{z}}{2} \\
  H_{r} &=& \sum_{k}\omega_{k}a_{k}^{\dagger}a_{k} \\
  H_{ar} &=& (\sigma_{+}+\sigma_{-})\sum_{k}g_{k}\left(a_{k}^{\dagger}+a_{k}\right)
\end{eqnarray}
where $\omega_{0}$ is the atomic transition frequency between the
ground state $|0\rangle$ and excited state $|1\rangle$.
$\sigma_{z}=|1\rangle\langle1|-|0\rangle\langle0|$,
$\sigma_{+}=|1\rangle\langle0|$ and $\sigma_{-}=|0\rangle\langle1|$
are pseudo-spin operators of atom. The index $k$ labels the field
modes of the reservoir with frequency $\omega_{k}$,
$a_{k}^{\dagger}$ and $a_{k}$ are the modes' creation and
annihilation operators, and $g_{k}$ is the frequency-dependent
coupling constant between the transition $|1\rangle-|0\rangle$ and
the field mode
 $k$.
\par The reduced non-perturbative non-Markovian quantum master equation of atom
 could be obtained by path integral\cite{Ish}
\begin{equation}\label{e5}
    \frac{\partial}{\partial
    t}\rho_{a}=-i\mathfrak{L}_{a}\rho_{a}-\int_{0}^{t}ds\langle \mathfrak{L}_{ar}e^{-i\mathfrak{L}_0(t-s)}\mathfrak{L}_{ar}e^{-i\mathfrak{L}_{0}(s-t)} \rangle_{r}\rho_{a}
\end{equation}
where $\mathfrak{L}_0$, $\mathfrak{L}_a$ and $\mathfrak{L}_{ar}$ are
Liouvillian operators defined as
\begin{eqnarray*}
  \mathfrak{L}_{0}\rho &\equiv& [H_{a}+H_{r},\rho] \\
  \mathfrak{L}_{a}\rho &\equiv& [H_{a},\rho] \\
  \mathfrak{L}_{ar}\rho &\equiv& [H_{ar},\rho]
\end{eqnarray*}
and $\langle...\rangle_{r}$ stands for partial trace of the
reservoir.
\par Then, we assume the reservoir is initially in squeezed vacuum
state\cite{scu}
\begin{equation}\label{n1}
    \rho_{r}=\prod_{k}S_{k}|0_{k}\rangle\langle0_{k}|S^{\dagger}_{k}
\end{equation}

\begin{equation}
    S_{k}=exp(re^{-i\theta}a_{k_{c}+k}a_{k_{c}-k}-re^{i\theta}a^{\dagger}_{k_{c}+k}a^{\dagger}_{k_{c}-k})
\end{equation}
$r$, $\theta$ are squeeze parameters, $|0_{k}\rangle$ is the vacuum
state of mode $k$. The central frequency of squeezing device is
$\omega_{0}=ck_{c}$, which corresponds to multimode squeezed vacuum
state with the central frequency equal to the cavity resonance
frequency and atom transition frequency. And the spectral density of
the reservoir is in Lorentzian form\cite{bel,sca}

\begin{eqnarray}\label{ee1}
  J(\omega) &=& \sum_{k}g^{2}_{k}\{\delta(\omega-\omega_{k})+\delta(\omega+\omega_{k})\} \nonumber \\
   &=& \frac{1}{2 \pi} \frac{\lambda \gamma^{2}}{(\omega-\omega_{0})^{2}+\gamma^{2}}
\end{eqnarray}
 where $\gamma$ represents the width of the spectral distribution of the
reservoir modes and is related to the correlation time of the noise
induced by the reservoir, $\tau_{r}=1/\gamma$. The parameter
$\lambda$ is related to the subsystem-reservoir  coupling strength.
There are two correlation functions in this model\cite{ccq}
\begin{eqnarray}\label{big1}
  \alpha_{1} &=& \int_{-\infty}^{\infty}J(\omega)e^{-i(\omega-\omega_{0})t}=\frac{\gamma\lambda}{2}e^{-\gamma t} \\
  \alpha_{2} &=& \int_{-\infty}^{\infty}J(\omega)e^{i(\omega+\omega_{0})t}=\frac{\gamma\lambda}{2}e^{(-\gamma+i2\omega_{0} )t}
\end{eqnarray}
$\alpha_{1}$ comes from the rotating-wave interaction and
$\alpha_{2}$ from the counter-rotating wave interaction.
\par So, the non-perturbative master equation of the
subsystem could be derived from Eq.(\ref{e5})
\begin{eqnarray}\label{e7}
  \frac{\partial}{\partial
    t}\rho_{a} &=& -\Gamma\rho_{a}+[\varepsilon_{0}J_{0}+\varepsilon_{+}J_{+}+ \varepsilon_{-}J_{-}]\rho_{a} \nonumber\\
    & & +[\upsilon_{0}K_{0}+\upsilon_{+}K_{+}+ \upsilon_{-}K_{-}]\rho_{a}
\end{eqnarray}
where $J_{0}$, $J_{+}$, $J_{-}$, $K_{0}$, $K_{+}$ and $K_{-}$ are
superoperators defined as
\begin{eqnarray*}
  J_{0}\rho_{a} &\equiv& \left[\frac{\sigma_{z}}{4},\rho_{a}\right] \\
  J_{+}\rho_{a} &\equiv& \sigma_{+}\rho_{a}\sigma_{+} \\
  J_{-}\rho_{a} &\equiv&  \sigma_{-}\rho_{a}\sigma_{-} \\
  K_{0}\rho_{a} &\equiv& (\sigma_{+}\sigma_{-}\rho_{a}+\rho_{a}\sigma_{+}\sigma_{-}-\rho_{a})/2 \\
  K_{+}\rho_{a} &\equiv& \sigma_{+}\rho_{a}\sigma_{-} \\
  K_{-}\rho_{a} &\equiv& \sigma_{-}\rho_{a}\sigma_{+}
\end{eqnarray*}
and
\begin{eqnarray*}
  \Gamma &=& 2M^{R}f+2(M^{R}\alpha^{R}+M^{I}\alpha^{I})+(2N+1)(f+\alpha^{R}) \\
  \varepsilon_{0} &=& -i2[\omega_{0}+2(M^{I}\alpha^{R}-M^{I}f-M^{R}\alpha^{R})-(2N+1)\alpha^{R}] \\
  \varepsilon_{+} &=& 2Mf+2M^{*}\alpha+(2N+1)(f+\alpha) \\
  \varepsilon_{-} &=& 2M\alpha^{*}+2M^{*}f+(2N+1)(f+\alpha^{*})
\end{eqnarray*}
\begin{eqnarray*}
  \upsilon_{0} &=& 2(\alpha^{R}-f) \\
  \upsilon_{+} &=& 2[M^{R}f+M^{R}\alpha^{R}+M^{I}\alpha^{I}+Nf+(N+1)\alpha^{R}]\\
  \upsilon_{-} &=& 2[M^{R}f+M^{R}\alpha^{R}+M^{I}\alpha^{I}+(N+1)f+N\alpha^{R}]
\end{eqnarray*}
\begin{eqnarray*}
 \alpha &=& \frac{\lambda\gamma(1-e^{-(\gamma+i2\omega_{0})t})}{2(\gamma+i2\omega_{0})} \\
  f &=& \frac{\lambda}{2}[1-exp(-\gamma t)]
\end{eqnarray*}
$N=sinh^{2}r$, $M=-e^{i\theta}sinhrcoshr$.  $\alpha^{R}$,
$\alpha^{I}$, $\alpha^{*}$ and $M^{R}$, $M^{I}$, $M^{*}$ are real
part, image part and conjugate of $\alpha$ and $M$, respectively.
Here, $N$ is the average photon number of every mode in squeezed
vacuum cavity. And $M$, $M^{*}$ represent the phase-dependent
correlation between different modes as $\langle b_{k}b_{k'}\rangle$
and $\langle b^{\dagger}_{k}b^{\dagger}_{k'}\rangle$,
respectively\cite{scu}.

\subsection{ Exact solution of master equation}
\par The time evolution of density operator in Eq.(\ref{e7}) could be obtained with
algebraic approach in Ref.\cite{zhang} because the superoperators
herein satisfy $SU(2)$ Lie algebraic communication relations, i.e.
\begin{eqnarray}
  \left[J_{-},J_{+}\right]\rho_{a} &=& -2J_{0}\rho_{a} \nonumber\\
  \left[J_{0},J_{\pm}\right]\rho_{a} &=& \pm J_{\pm} \rho_{a} \nonumber\\
  \left[K_{-},K_{+}\right]\rho_{a} &=& -2K_{0}\rho_{a} \nonumber\\
  \left[K_{0},K_{\pm}\right]\rho_{a} &=& \pm K_{\pm}\rho_{a} \nonumber\\
  \left[K_{i},J_{j}\right] &=& 0
\end{eqnarray}
where $i,j=0, \pm$. By directly integrating Eq.(\ref{e7}), the
formal solution is obtained as\cite{pur}
\begin{eqnarray}\label{e8}
  \rho_{a}(t) &=& e^{-\Gamma_{k}}\hat{T}e^{\int_{0}^{t}dt(\varepsilon_{0}J_{0}+ \varepsilon_{+}J_{+}+\varepsilon_{-}J_{-})}\nonumber\\
    & & \times \hat{T}e^{\int_{0}^{t}dt(\nu_{0}K_{0}+\nu_{+}K_{+}+\nu_{-}K_{-})}\rho_{a}(0)
\end{eqnarray}
where $\hat{T}$ is time ordering operator and
\begin{eqnarray*}
 \Gamma_{k} &=& (2M^{R}+2N+1)(F+\tilde{\alpha}^{R})+2M^{I}\tilde{\alpha}^{I}) \\
  \tilde{\alpha} &=& \int_{0}^{t}\alpha dt=\frac{\lambda\gamma(t-\frac{1-e^{-(\gamma+i2\omega_{0})t}}{\gamma+i2\omega_{0}})}{2(\gamma+i2\omega_{0})}\equiv\tilde{\alpha}^{R}+i\tilde{\alpha}^{I} \\
  \tilde{\alpha}^{*} &=& \tilde{\alpha}^{R}-i\tilde{\alpha}^{I} \\
  F(t) &=& \lambda\{t-[1-exp(-\gamma t)]/\gamma\}/2
\end{eqnarray*}
where $\tilde{\alpha}^{R}$, $\tilde{\alpha}^{I}$ and
$\tilde{\alpha}^{*}$ are real part, image part and conjugate of
$\tilde{\alpha}$, respectively.

\par The exponential functions of superoperators in Eq.(\ref{e8}) could be disentangled in
form\cite{pur}
\begin{eqnarray}
  \hat{T}e^{\int_{0}^{t}dt(\varepsilon_{0}J_{0}+ \varepsilon_{+}J_{+}+\varepsilon_{-}J_{-})} &=& e^{j_{+}J_{+}}e^{j_{0}J_{0}}e^{j_{-}J_{-}} \\
  \hat{T}e^{\int_{0}^{t}dt(\nu_{0}K_{0}+\nu_{+}K_{+}+\nu_{-}K_{-})} &=& e^{k_{+}K_{+}}e^{k_{0}K_{0}}e^{k_{-}K_{-}}
\end{eqnarray}
where $j_{+}$, $j_{0}$, $j_{-}$ and $k_{+}$, $k_{0}$, $k_{-}$
satisfy the following differential equation
\begin{eqnarray}
  \dot{X}_{+} &=& \mu_{+}-\mu_{-}X_{+}^{2}+\mu_{0}X_{+} \\
  \dot{X}_{0} &=& \mu_{0}-2\mu_{-}X_{+}\\
  \dot{X}_{-} &=& \mu_{-}exp(X_{0})
\end{eqnarray}
$\mu=\varepsilon$ for $X=j$ and $\mu=\nu$ for $X=k$.

\par Using the results above, the exact solution of the master
equation Eq.(\ref{e7}) is obtained
\begin{equation}\label{e19}
     \rho_{a}(t) = e^{-\Gamma_{k}}\tilde{\rho}(t)
\end{equation}
\begin{equation}\label{e20}
    \tilde{\rho}(t)=\left(\begin{array}{cc}
                           l\rho^{11}_{a}(0)+ m\rho^{00}_{a}(0) & x\rho^{10}_{a}(0)+ y\rho^{01}_{a}(0) \\
                           q\rho^{01}_{a}(0)+ r\rho^{10}_{a}(0) &
                           n\rho^{00}_{a}(0)+ p\rho^{11}_{a}(0)
                         \end{array}
                         \right)
\end{equation}
\begin{eqnarray}
  l &=& e^{k_{0}/2}+e^{-k_{0}/2}k_{+}k_{-},\ m = e^{-k_{0}/2}k_{+} \\
  n &=& e^{-k_{0}/2},\ p= e^{-k_{0}/2}k_{-}\\
  q &=& e^{-j_{0}/2},\ r= e^{-j_{0}/2}j_{-} \\
  x &=& e^{j_{0}/2}+e^{-j_{0}/2}j_{+}j_{-},\ y=e^{-j_{0}/2}j_{+}
\end{eqnarray}

\subsection{ Concurrence}
\par In order to investigate the entanglement dynamics of the bipartite
system, we use Wootters concurrence\cite{woott}. For simplicity, we
assume that the two subsystems have the same parameters. The
concurrence of the whole system could  be obtained\cite{ikr}
\begin{eqnarray}
  C_{\xi} &=& max \left\{0,c_{1},c_{2}\right\},(\xi=\Phi,\Psi) \\
  c_{1} &=& 2e^{-2\Gamma_{k}}(\sqrt{\rho_{23}\rho_{32}}-\sqrt{\rho_{11}\rho_{44}}) \nonumber\\
  c_{2} &=&
  2e^{-2\Gamma_{k}}(\sqrt{\rho_{14}\rho_{41}}-\sqrt{\rho_{22}\rho_{33}})\nonumber
\end{eqnarray}
corresponding to the initial states of $|\Phi\rangle =
\beta|01\rangle+\eta|10\rangle$ and $|\Psi\rangle =
\beta|00\rangle+\eta|11\rangle$, respectively. Where $\beta$ is real
and $0<\beta<1$, $\eta=|\eta|e^{i\varphi}$ and
$\beta^{2}+|\eta|^{2}=1$. For maximum entanglement Bell state,
$\beta$ is equal to $\sqrt{2}/2$. The reduced joint density matrix
of the two atoms, in the standard product basis
$\mathfrak{B}=\{|1\rangle\equiv|11\rangle,
|2\rangle\equiv|10\rangle, |3\rangle\equiv|01\rangle,
|4\rangle\equiv|00\rangle \}$, could be obtained by the method in
ref.\cite{bel}
\begin{equation}\label{e28}
    \rho^{AB}=e^{-2\Gamma_{k}}
    \left(\begin{array}{cccc}
            \rho_{11} & 0 & 0 & \rho_{14} \\
            0 & \rho_{22} & \rho_{23} & 0 \\
            0 & \rho_{32} & \rho_{33} & 0 \\
            \rho_{41} & 0 & 0 & \rho_{44}
          \end{array}
          \right)
\end{equation}
here the diagonal elements are
\begin{eqnarray*}
  \rho_{11} &=& l^{2}\rho_{11}(0)+lm\rho_{22}(0)+ml\rho_{33}(0)+m^{2}\rho_{44}(0) \\
  \rho_{22} &=& lp\rho_{11}(0)+lm\rho_{22}(0)+mp\rho_{33}(0)+mn\rho_{44}(0) \\
  \rho_{33} &=& lp\rho_{11}(0)+pm\rho_{22}(0)+nl\rho_{33}(0)+nm\rho_{44}(0) \\
  \rho_{44} &=& p^{2}\rho_{11}(0)+pn\rho_{22}(0)+np\rho_{33}(0)+n^{2}\rho_{44}(0)
\end{eqnarray*}
and the nondiagonal elements are
\begin{eqnarray*}
  \rho_{14} &=& x^{2}\rho_{14}(0)+xy\rho_{23}(0)+yx\rho_{32}(0)+y^{2}\rho_{41}(0) \\
  \rho_{23} &=& xr\rho_{14}(0)+xq\rho_{23}(0)+yr\rho_{32}(0)+yq\rho_{41}(0) \\
  \rho_{32} &=& rx\rho_{14}(0)+ry\rho_{23}(0)+qx\rho_{32}(0)+qy\rho_{41}(0) \\
  \rho_{41} &=& r^{2}\rho_{14}(0)+rq\rho_{23}(0)+qr\rho_{32}(0)+q^{2}\rho_{41}(0)
\end{eqnarray*}

 \section{\label{sec:num1} Numerical results and discussion}

\par In order to study the effects of non-Markovian squeezed reservoir
 on the decoherence, we assume that $\lambda$ is principally equal to $10\gamma$ in Eq.(\ref{ee1}), which could
 be realized in a high-Q cavity\cite{bel}.
 \par First, we focus on the effects of squeeze parameters on the decoherence of
 two qubits with initial maximum Bell entanglement states $|\Phi\rangle$ and $|\Psi\rangle$, respectively.
\par (A) For $\omega_{0}=10\gamma$ and $\beta=\sqrt{2}/2$,
Fig.1 and  Fig.2 show that, for the case of initial maximum Bell
state $|\Phi\rangle$, the concurrence first decreases to a certain
value and then revives before it vanishes, while that periodically
vanishes and revives with a damping of its revival amplitude for the
case of initial maximum Bell state $|\Psi\rangle$. It also reveals
that the amplitude and the duration time of entanglement revival are
different for the case of $\pi/2\leq\theta\leq3\pi/2$ and for the
case of $0\leq\theta\leq\pi/2$ and $3\pi/2\leq\theta\leq2\pi$.

\begin{figure}
  \includegraphics{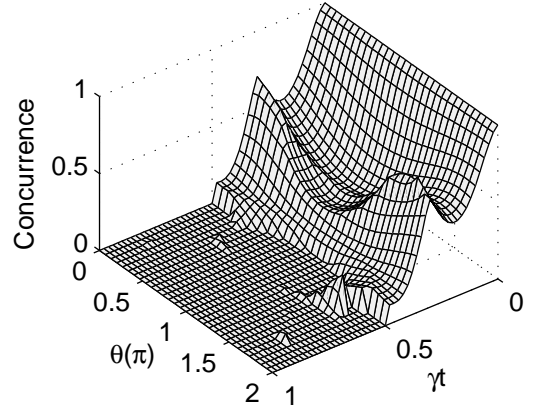}\\
  \caption{Concurrence $C_{\Phi}$ as a function of $\gamma t$ and $\theta$ with $\lambda=10\gamma$, $\omega_{0}=10\gamma$, $\beta=\sqrt{2}/2$ and $r=0.2$.}\label{F1}
\end{figure}
\begin{figure}
  \includegraphics{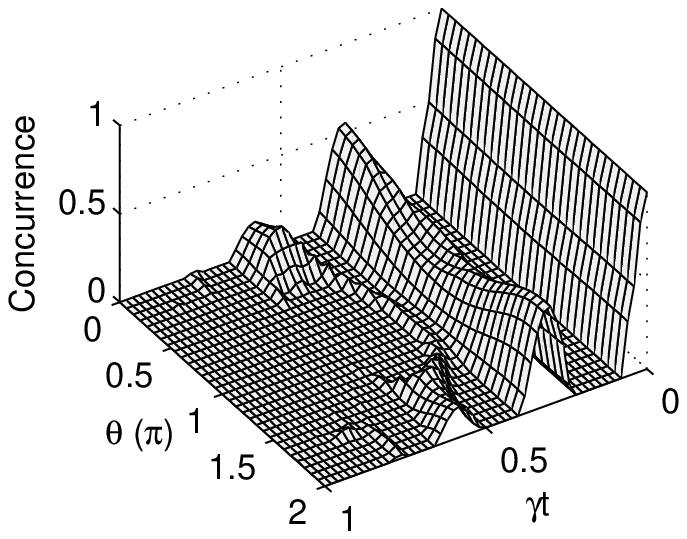}\\
  \caption{Concurrence $C_{\Psi}$ as a function of $\gamma t$ and $\beta^{2}$ with $\lambda=10\gamma$, $\omega_{0}=10\gamma$, $\beta=\sqrt{2}/2$ and $r=0.2$.}\label{F2}
\end{figure}

\begin{figure}
  \includegraphics{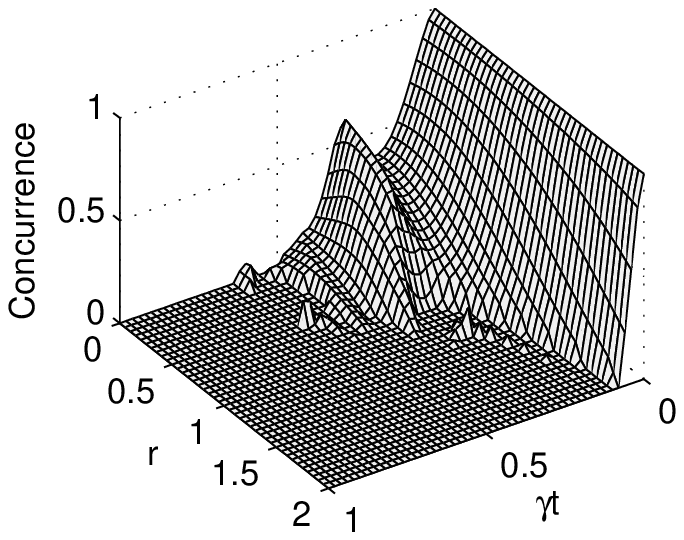}\\
  \caption{Concurrence $C_{\Phi}$ as a function of $\gamma t$ and $r$ with $\lambda=10\gamma$, $\omega_{0}=10\gamma$, $\beta=\sqrt{2}/2$ and $\theta=\pi/4$.}\label{F3}
\end{figure}

\begin{figure}
  \includegraphics{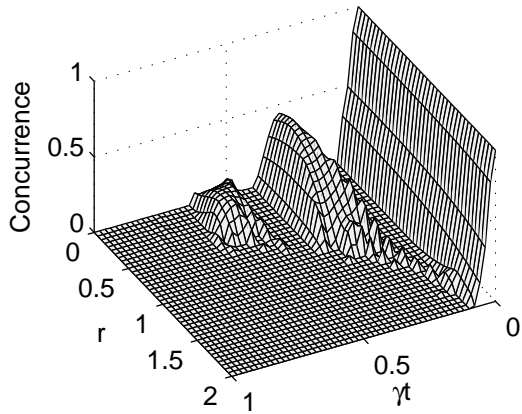}\\
  \caption{Concurrence $C_{\Psi}$ as a function of $\gamma t$ and $r$ with $\lambda=10\gamma$, $\omega_{0}=10\gamma$, $\beta=\sqrt{2}/2$ and $\theta=\pi/4$.}\label{F4}
\end{figure}

\par (B)For $\omega_{0}=10\gamma$ and $\beta=\sqrt{2}/2$,
Fig.3 and Fig.4 show that, for $r\leq1$, the concurrence first
decreases to a certain value and then revives before it vanishes for
the case of initial maximum Bell state $|\Phi\rangle$, while that
periodically vanishes and revives with damping amplitude for the
case of initial maximum Bell state $|\Psi\rangle$. And the
concurrence decreases monotonically and vanishes permanently after a
short time for $r>1$. The results exhibit that the disentanglement
time decreases with the increasing of squeeze parameter $r$ for
fixed $\theta$, due to the increasing of average photon number of
every mode in squeezed vacuum cavity.

\par The above results also reveal that the decoherence behavior of
concurrence is sensitive to squeeze parameters and insensitive to
initial state.

\begin{figure}
  \includegraphics{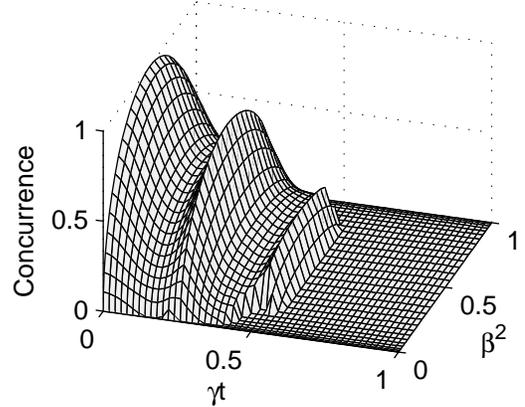}\\
  \caption{Concurrence $C_{\Phi}$ as a function of $\gamma t$ and $\beta^{2}$ with $\lambda=10\gamma$, $\omega_{0}=12\gamma$, $\theta=\pi/4$ and $r=0.2$.}\label{F5}
\end{figure}

\begin{figure}
  \includegraphics{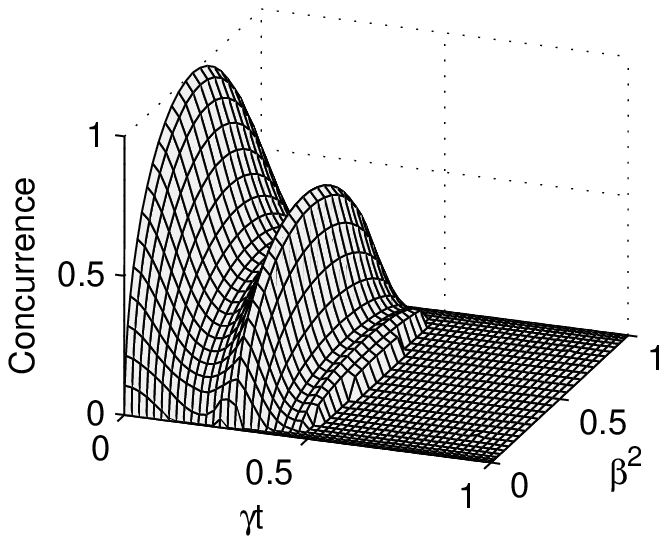}\\
  \caption{Concurrence $C_{\Phi}$ as a function of $\gamma t$ and $\beta^{2}$ with $\lambda=10\gamma$, $\omega_{0}=10\gamma$, $\theta=\pi/4$ and $r=0.2$.}\label{F6}
\end{figure}

\begin{figure}
  \includegraphics{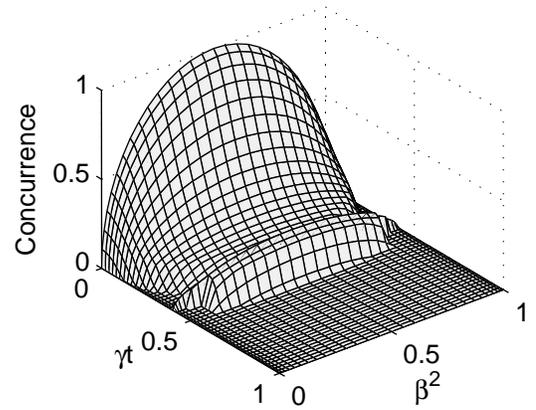}\\
  \caption{Concurrence $C_{\Phi}$ as a function of $\gamma t$ and $\beta^{2}$ with $\lambda=10\gamma$, $\omega_{0}=6.5\gamma$, $\theta=\pi/4$ and $r=0.2$.}\label{F7}
\end{figure}

\par Then, for fixed squeeze parameters
$\theta=\pi/4$ and $r=0.2$, the decoherence for different initial
states was discussed.

\par (A) The decoherence behavior of $C_{\Psi}$ is discussed as follows.

\par (1) From Fig.5, Fig.6 and Fig.7, we could find that the concurrence $C_{\Phi}$
will decreases and revives before it vanishes permanently. With the
ratio of the coupling strength to the atomic frequency increasing,
the times of entanglement revival decreases because of the
increasing of the average value of correlation function
$\alpha_{2}$, corresponding to the counter-rotating wave terms.

\par (2) For $\omega_{0}=3\gamma$ and $\lambda=20\gamma$, Fig.8 exhibits that the
concurrence $C_{\Phi}$ decreases monotonically and vanishes
permanently in a short time even in a non-Markovian squeezed
reservoir, resulted from the strong interaction between atom and the
non-Markovian squeezed reservoir.

\begin{figure}
  \includegraphics{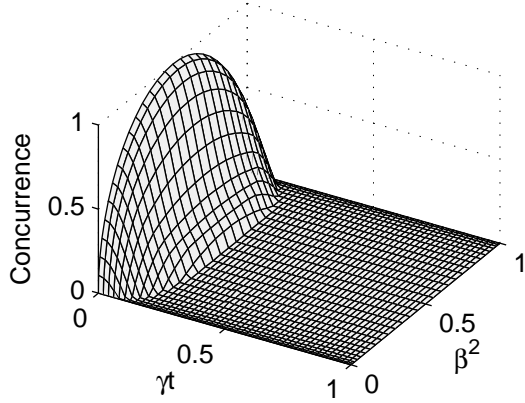}\\
  \caption{Concurrence $C_{\Phi}$ as a function of $\gamma t$ and $\beta^{2}$ with $\lambda=10\gamma$, $\omega_{0}=3\gamma$, $\theta=\pi/4$ and $r=0.2$.}\label{F8}
\end{figure}

\begin{figure}
  \includegraphics{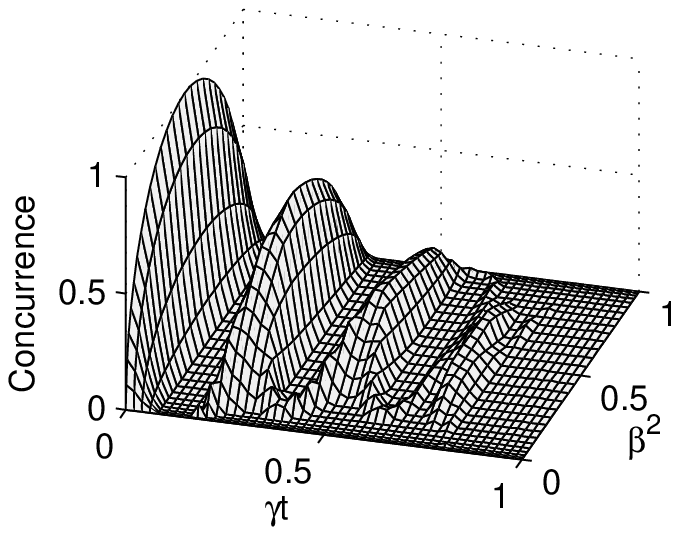}\\
  \caption{Concurrence $C_{\Psi}$ as a function of $\gamma t$ and $\beta^{2}$ with $\lambda=10\gamma$, $\omega_{0}=12\gamma$, $\theta=\pi/4$ and $r=0.2$.}\label{F9}
\end{figure}

\begin{figure}
  \includegraphics{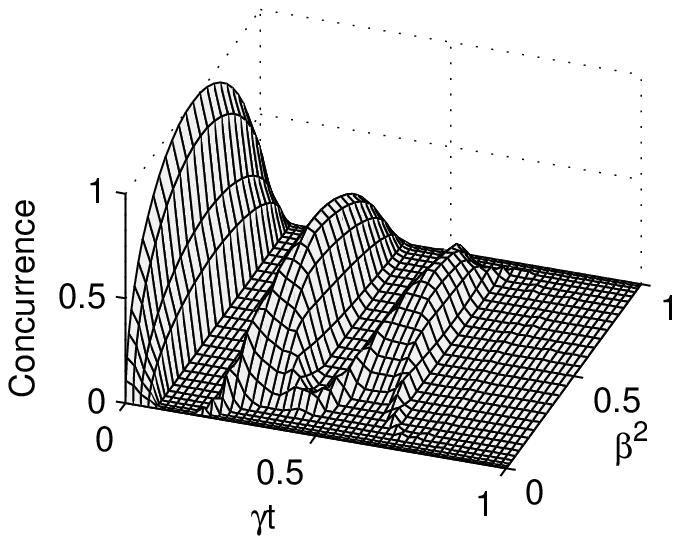}\\
  \caption{Concurrence $C_{\Psi}$ as a function of $\gamma t$ and $\beta^{2}$ with $\lambda=10\gamma$, $\omega_{0}=10\gamma$, $\theta=\pi/4$ and $r=0.2$.}\label{F10}
\end{figure}

\begin{figure}
  \includegraphics{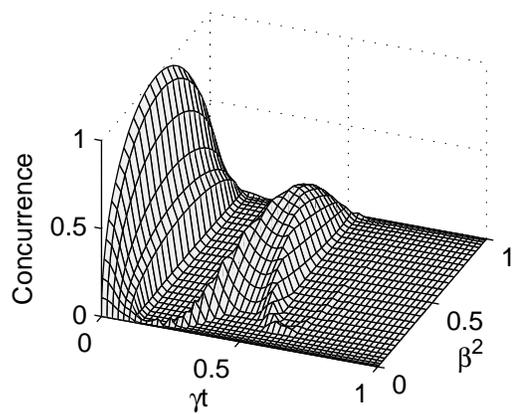}\\
  \caption{Concurrence $C_{\Psi}$ as a function of $\gamma t$ and $\beta^{2}$ with $\lambda=10\gamma$, $\omega_{0}=6.5\gamma$, $\theta=\pi/4$ and $r=0.2$.}\label{F11}
\end{figure}

\par (B)The decoherence behavior of $C_{\Psi}$ is discussed as follows.

\par (1) From Fig.9 , Fig.10 and Fig.11, we find that the concurrence periodically
vanishes and revives with a damping of its revival amplitude. With
the ratio of the coupling strength to the atomic frequency
increasing, the times of entanglement revival decrease. There are
entanglement revivals after a period of time of disentanglement,
which are different from that of the case of $C_{\Phi}$.

\par (2) For $\omega_{0}=2\gamma$ and $\lambda=20\gamma$, the evolution dynamics
of concurrence $C_{\Psi}$ is almost the same as that in Fig.8.
Unlike the two cases above, the evolution behavior of concurrence
$C_{\Psi}$ becomes symmetric because the strong coupling of atom
with non-Markovian reservoir and the effect of counter-rotating wave
interaction.

\par The above results reveal that, in principal, the decoherence of $C_{\Phi}$ is
symmetrical with $\beta^{2}$ because of the symmetry of initial
state $|\Phi\rangle$, while that is unsymmetrical with $\beta^{2}$
because the initial state $|\Psi\rangle$ is unsymmetrical with
$\beta^{2}$. However, the strong coupling and the counter-rotating
wave interaction could make the decoherence of $C_{\Psi}$ become
symmetrical with $\beta^{2}$. With the enhancing of coupling
strength, the memory effect of the counter-rotating wave terms
becomes dominant because the the average value of correlation
function $\alpha_{2}$ increases. The results are quite different
from that of the case for qubits in a vacuum cavity\cite{bel}.

\section{\label{sec:clu}Conclusion}
\par The reduced non-perturbative quantum master equation of atom in
initially squeezed vacuum cavity has been derived and its exact
solution is obtained. The decoherence behaviors of two qubits with
squeeze parameters and other parameters have been discussed.
\par The results show that the decoherence behavior of two qubits in
a squeezed reservoir is dependent on the squeeze parameter, the
ratio of the coupling strength to the atomic transition frequency
and the ratio of the width of reservoir spectral density to the
atomic transition frequency. First, in principle, the
disentanglement time decreases with the increasing of squeeze
parameter $r$, due to the increasing of average photon number of
every mode in squeezed vacuum cavity. Second, there are entanglement
revivals after complete disentanglement for the case of even parity
initial Bell state, while there are entanglement decrease and
revival before complete disentanglement for the case of odd parity
initial Bell state. Third, with the enhancing of coupling strength,
the times of entanglement revival decrease due to the effect of
counter-rotating wave terms. Fourth, there is entanglement revival
due to the memory effect of the non-Markovian squeezed environment.

\begin{acknowledgments}
This work was supported by the National Natural Science Foundation
of China Grants No.60578055, the State Key Program for Basic
Research of China under Grant No. 2007CB925204 and No.2007CB307001.
\end{acknowledgments}

\end{document}